\documentclass[aps,twocolumn,prl,superscriptaddress,amsmath,showpacs]{revtex4}
\usepackage{epsfig,graphicx,times}
\usepackage{amstext}
\usepackage{amsmath}
\usepackage{amssymb}
\usepackage{graphicx}
\usepackage{latexsym}
\usepackage{bm}

\begin{document}
\title{Quantum  tomography for solid state qubits}

\author{ Yu-xi Liu}
\affiliation{Frontier Research System,  The Institute of Physical
and Chemical Research (RIKEN), Wako-shi 351-0198}
\author{L.F. Wei}
\affiliation{Frontier Research System,  The Institute of Physical
and Chemical Research (RIKEN), Wako-shi 351-0198}
\affiliation{Institute of Quantum Optics and Quantum Information,
Department of Physics, Shanghai Jiaotong University, Shanghai
200030, P.R. China }
\author{Franco Nori}
\affiliation{Frontier Research System,  The Institute of Physical
and Chemical Research (RIKEN), Wako-shi 351-0198}
\affiliation{Center for Theoretical Physics, Physics Department,
Center for the Study of Complex Systems, The University of
Michigan, Ann Arbor, Michigan 48109-1120}

\date{\today}

\begin{abstract}
We propose a  method for the tomographic reconstruction of qubit
states for a general class of  solid state systems in which the
Hamiltonians are represented by spin operators, e.g., with
Heisenberg-, $XXZ$-, or $XY$- type exchange interactions. We
analyze the implementation of the projective operator
measurements, or spin measurements, on qubit states. All  the
qubit states for the spin Hamiltonians can be reconstructed by
using experimental data.

\pacs{03.65.Ta, 03.65.Wj}
\end{abstract}

\maketitle \pagenumbering{arabic}

Quantum information processing requires the effective measurement
of quantum states. However, a single quantum measurement can only
obtain partial information of a quantum state. The reconstruction
of a quantum state requires  measuring a complete set of
observables on an ensemble of identically prepared copies of the
system. This method called quantum state tomography~\cite{band},
is very important because any unknown state can be ascertained by
tomographic measurements. Moreover, the full description of qubit
states can increase the accuracy of quantum operations.
Tomographic measurements have been experimentally implemented for,
e.g., the nuclear spin state of an NMR system~\cite{chuang}, the
electromagnetic field and photon state~\cite{sm}, the vibration
state of molecules~\cite{dunn}, the motional quantum state of a
trapped atom~\cite{lei},  and atomic wave packets~\cite{kur}.

Experimental investigations on solid state qubits are very
promising, especially in superconducting~\cite{moo,yu} and quantum
dot structures~\cite{chen}. These recent achievements make it
necessary to experimentally determine  quantum states in solid
state systems. Although there are many theoretical studies on
tomography (e.g., references~\cite{theor} and references therein),
to our knowledge, these are not specific to solid state systems.
Here, we focus on this question for quantum computing models using
standard spin representations for solid state qubits. Our proposal
is related to tomographic measurements using NMR. The measurements
of the density matrix in NMR experiments are obtained from the NMR
spectrum of the linear combination of ``product operators", i.e.
products of the  angular momentum operators~\cite{cory}. However,
experiments in  solid state systems usually involve the local
single-qubit projective operator measurement (POM) or spin
measurement. So, we study the method of tomographic reconstruction
of solid state qubits by  POM or spin measurement for a number of
promising solid-state quantum computing
models~\cite{makhlin,as,loss,ww,kane,dm,zorin,you}. We will
investigate how  the multi-qubit correlation measurements can be
realized by virtue of an appropriate two-qubit operation combined
with single qubit operations.

{\it State and measurements.} --- Using the density matrix form,
an $n$-qubit state $\rho$ can be expressed as
\begin{equation}\label{eq:1}
\rho=\frac{1}{2^n}\sum_{l_{1},\cdots,l_{n}=0,x,y,z}r_{l_{1}\cdots
l_{n}}\,\,\sigma_{l_{1}}\otimes
\sigma_{l_{2}}\otimes\cdots\otimes\sigma_{l_{n}},
\end{equation}
where $r_{l_{1}\cdots l_{n}}$ are $4^{n}$ real parameters,
$\sigma_{l_{m}=x,y,z} $ and $\sigma_{l_{m}=0} \,\,(0\leq m\leq n)$
are the Pauli spin and identity operators of the $m$th qubit
respectively. We adopt the convention
$|0\rangle=|\!\uparrow\rangle$ and
$|1\rangle=|\!\downarrow\rangle$ to denote the computational basis
states of each qubit. The normalization condition ${\rm
Tr}{\rho}=1$ makes $r_{0,\cdots,0}=1$ which means that  $\rho$ can
be specified by $(4^n-1)$ real parameters. These parameters
correspond to the expectation values of the measurements given by
the operators $\sigma_{j_{1}}\otimes\cdots\otimes\sigma_{j_{n}}$;
that is, ${\rm
Tr}\{\rho(\sigma_{j_{1}}\otimes\cdots\otimes\sigma_{j_{n}})\}
=r_{l_{1}\cdots
l_{n}}\delta_{j_{1}l_{1}}\cdots\delta_{j_{n}l_{n}}$, where
$l_{1},\cdots, l_{n}$ are not simultaneously taken as zero. If
there are $n-m$ identity operators among
$\sigma_{l_{1}}\otimes\cdots\otimes\sigma_{l_{n}}$, the
measurement is really done by the $m$ qubits and it can be
abbreviated by the tensor product of only the $m$ Pauli operators,
which is denoted hereafter by
$\sigma_{1\,l_{1}}\otimes\cdots\otimes\sigma_{m\,l_{m}}$. The
$(4^n-1)$ measurements required to reconstruct the $n$-qubit state
can be decomposed into a summation from the single-qubit to
$n$-qubit measurements as $\sum_{j=1}^{n}3^{j}\binom{n}{j}$; where
$3^{j}\binom{n}{j}$ is the number of $j$-qubit measurements and
$\binom{n}{j}$ is the binomial coefficient.

To reconstruct the $n$-qubit state $\rho$, we need to determine
all of its expanded coefficients $\{r_{l_{1},\cdots,l_{n}}\}$. In
solid state systems, the correlated multi-qubit measurement is not
realizable now, and the experimental readout is often done via
single-qubit POM (e.g., Refs.~\cite{loss,makhlin,ww,as}) or single
spin measurement (e.g., Refs.~\cite{zorin,you}).  Without loss of
generality, we assume that the POM is denoted by $|1\rangle\langle
1|$ and the spin measurement is presented by $\sigma_{x}$ or
$\sigma_{y}$. Below, we will discuss how to determine the
$(4^n-1)$ coefficients of an $n$-qubit state by using experimental
data of the POM and then generalize it to the spin case. Our goal
is to build a correspondence between above the measurements and
actual measurements done via $|1\rangle\langle 1|$.

{\it Single-qubit measurements.}--- The single-qubit measurements
$\sigma_{lz}\,\,(l=1,2,\cdots, n)$ can be done by the projectors
$(|1\rangle\langle 1|)_{l}$ due to the equivalence
$\sigma_{lz}=\sigma_{l0}-2(|1\rangle\langle 1|)_{l}$,  with
identity operators $\sigma_{l0}$. Thus the POM experiments can
directly determine $n$ coefficients via $n$ outcomes of the
measurements $(|1\rangle\langle 1|)_{l}$. The measurements
corresponding to the remaining $(4^n-n-1)$ coefficients cannot be
directly performed because of limitations of  current experiment.
In order to obtain these coefficients, a sequence of quantum
operations $W$ is needed such that each coefficient can be
transformed to a position that is measurable by a POM experiment.
The probability $p$ of the $l$th single-qubit measurement
$(|1\rangle\langle 1|)_{l}$ on the operated state $\rho$ can be
expressed as
\begin{equation}
p={\rm Tr}\{W\rho W^{\dagger}(|1\rangle\langle 1|)_{l}\} ={\rm
Tr}\{\rho W^{\dagger}(|1\rangle\langle 1|)_{l}W\},
\end{equation}
which means that the experimental POM $(|1\rangle\langle 1|)_{l}$
on the state $W\rho W^{\dagger}$ can be equivalently described as
$W^{\dagger}(|1\rangle\langle 1|)_{l}W=
\left[\frac{1}{2}-\frac{1}{2}W^{\dagger}\sigma_{lz}W\right]$ on
the original state $\rho$. Because the POM experiment is
equivalent to $W^{\dagger}\sigma_{lz}W$, we can call it an
equivalent $m$-qubit measurement of $(|1\rangle\langle 1|)_{l}$,
if the largest number of  Pauli operators in the operator product
 expansion of $W^{\dagger}\sigma_{lz}W$ is an
$m$-qubit. Thus the $(4^n-n-1)$ coefficients can be obtained by
the single-qubit projection $(|1\rangle\langle 1|)_{l}$ on the
state $\rho$ with appropriate operations $W$ implemented by the
dynamical evolution of the system with experimentally controllable
parameters.

For all universal quantum computing proposals, the most general
Hamiltonian of the system can be described as
\begin{equation}\label{eq:4}
H=\sum_{l=1}^{n}\sum_{\alpha=x,y,z}\varepsilon_{l\,\alpha}\,\sigma_{l\,\alpha}+
\sum_{1=l<m}^{n}\,\,\sum_{\alpha,\beta=x,y,x}J_{l\,m}^{\alpha\beta}\,\,\sigma_{l\,\alpha}\otimes\sigma_{m\,\beta},
\end{equation}
where  $\{\varepsilon_{l\,\alpha}\}$ and
$\{J^{\alpha\beta}_{l\,m}\}$ are controllable and tunable
system-specific one-qubit and exchange coupling parameters, which
are required by the universality of quantum computing~\cite{dp},
$\sigma_{l\,\alpha=x,y,z}$ denote the Pauli operators of the $l$th
qubit. Without loss of generality, all parameters are assumed to
be positive real numbers.

In order to obtain each coefficient corresponding to single-qubit
measurements $\sigma_{lx}$ or $\sigma_{ly}$, all single-qubit
operations need to be performed  separately by controlling the
one-qubit parameters $\varepsilon_{l\,\alpha}$ while turning off
all interactions in the Hamiltonian~(\ref{eq:4}), that is,
$J^{\alpha\beta}_{l\,m}=0$. For $n$ single-qubit measurements
$\{\sigma_{ly}\}$, each $\sigma_{ly}$ can be equivalently obtained
by $(|1\rangle\langle 1|)_{l}$, after the $l$th qubit is rotated
$\pi/2$  about the $x$ axis, the latter expressed as
$X_{l}=\exp\{-i\pi \sigma_{lx}/4\}$. This rotation can be realized
within the evolution time $t=\hbar\pi/4\varepsilon_{lx}$, after
the one-qubit parameters $\varepsilon_{ly}$ and $\varepsilon_{lz}$
are adjusted to zero. Other $n$ single-qubit measurements
$\{\sigma_{lx} \}$ can also be obtained by measuring
$(|1\rangle\langle 1|)_{l}$ on the state,  within the evolution
time $t=\hbar\pi/4\varepsilon_{ly}$, after $\varepsilon_{lx}$ and
$\varepsilon_{lz}$ are set to zero. This  quantum operation is
equivalent to a $\pi/2$ rotation of the $l$th qubit about the $y$
axis, which is denoted by $Y_{l}=\exp\{-i\pi \sigma_{ly}/4\}$.
However, not all of the three one-qubit parameters
$\varepsilon_{lx}$, $\varepsilon_{ly}$ and $\varepsilon_{lz}$
appear in the Hamiltonian of most physical systems. For example,
the parameter $\varepsilon_{ly}$ is always zero in the  charge
qubit system~\cite{makhlin}. For this case, to obtain the rotation
angle $\theta$ about the $y$ axis  we need to alternatively  turn
on and off the single-qubit quantum operations: (1)
$\overline{X}_{l}=\exp\{-i\varepsilon_{lx}\sigma_{lx}t_{1}/\hbar\}$,
with the operation time $t_{1}=3\hbar\pi/4\varepsilon_{lx}$; (2)
$Z_{l}(\theta)=\exp\{-i\theta \sigma_{lz}/2\}$, with
$\theta=\varepsilon_{lz}t_{2}/\hbar$; and (3)
$X_{l}=\exp\{-i\varepsilon_{lx}\sigma_{lx}t_{3}/\hbar\}$,  with
$t_{3}=\hbar\pi/4\varepsilon_{lx}$. These can be expressed as
$Y_{l}(\theta)=X_{l}Z_{l}(\theta)\overline{X}_{l}=\exp\{-i\theta\sigma_{1y}/2\}$.
Especially, we denote the rotation $\pi/2$ about the $z$ axis by
$Z_{l}=\exp\{-i\pi\sigma_{lz}/4\}$.  In principle, if the $l$th
qubit system has only two controllable  one-qubit parameters
$\varepsilon_{\alpha}$ and $\varepsilon_{\beta}$, then  the
rotation angle $\epsilon_{\alpha\beta\gamma}\theta$  about the
axis $\gamma$ can be obtained by first doing a rotation of $\pi/4$
about the axis $\alpha$, then a rotation of $\theta$ about the
axis $\beta$, and finally a rotation by $-\pi/4$ about the axis
$\alpha$; that is,
$\exp\{-i\epsilon_{\alpha\beta\gamma}\theta\sigma_{\gamma}/2\}=e^{-i\pi\sigma_{\alpha}/4}
e^{-i\theta\sigma_{\beta}/2}e^{i\pi\sigma_{\alpha}/4}$,  where
$\alpha, \beta, \gamma$ can be $x, \,y$, or $z$ and the
Levi-Civita tensor $\epsilon_{\alpha\beta\gamma}$ is equal to $+1$
and $-1$ for the even and odd permutation of its indices,
respectively. To reconstruct a single-qubit state, three
single-qubit measurements $\sigma_{\alpha}\, (\alpha=x,\,y\,z)$
are sufficient to obtain $r_{z}$, which determines the
probabilities of finding $0$ and $1$, as well as $r_{x}$ and
$r_{y}$ which determine the relative phase of $|0\rangle$ and
$|1\rangle$.

{\it Two-qubit measurements.}---
\begingroup \squeezetable
\begin{table}
\caption{\label{tab1}Quantum operation and two-qubit
measurements.}
\begin{ruledtabular}
\begin{tabular}{c|c|c|c}
\multicolumn{0}{c}{$XY$ model} &  & \multicolumn{2}{c}{Heisenberg model} \\
 \cline{1-4}
EM& Operations & Equivalent measurement (EM) & Operations        \\
\hline
 $\sigma_{1y}+\sigma_{1x}\sigma_{1x}$ & $X_{1}U_{1}Y_{1}$ &
 $\sigma_{1z}+\sigma_{2z}+\sigma_{1y}\sigma_{2x}-\sigma_{1x}\sigma_{2y}$ & $U_{2}$ \\
\hline $-\sigma_{1z}+\sigma_{1x}\sigma_{2y}$& $ Y_{1}U_{1}Y_{1}$ &
 $\sigma_{1y}+\sigma_{2z}-\sigma_{1z}\sigma_{2x}-\sigma_{1x}\sigma_{2y}$ & $U_{2}X_{1}$     \\
\hline
 $-\sigma_{1z}-\sigma_{1x}\sigma_{2z}$& $Y_{1}U_{1}Y_{1}X_{2}$  &  $\sigma_{2z}-\sigma_{1x}
 +\sigma_{1y}\sigma_{2x}-\sigma_{1z}\sigma_{2y}$& $U_{2}Y_{1}$     \\
\hline
 $-\sigma_{1z}-\sigma_{1y}\sigma_{2x}$& $X_{1}U_{1}X_{1}$  &
  $\sigma_{1z}+\sigma_{2z}+\sigma_{1x}\sigma_{2x}+\sigma_{1y}\sigma_{2y}$ & $U_{2}Z_{1}$     \\
\hline
 $-\sigma_{1x}-\sigma_{1y}\sigma_{2y}$& $Y_{1}U_{1}X_{1}$  &
 $\sigma_{1z}-\sigma_{2x}+\sigma_{1y}\sigma_{2z}-\sigma_{1x}\sigma_{2y}$ & $U_{2}Y_{2}$   \\
\hline
 $-\sigma_{1x}+\sigma_{1y}\sigma_{2z}$& $Y_{1}U_{1}X_{1}X_{2}$  &
 $-\sigma_{1x}-\sigma_{2x}-\sigma_{1z}\sigma_{2y}+\sigma_{1y}\sigma_{2z}$   &  $Y_{1}U_{2}$  \\
\hline
 $\sigma_{1y}-\sigma_{1z}\sigma_{2x}$& $ X_{1}U_{1}$  &
 $\sigma_{1y}+\sigma_{2y}-\sigma_{1z}\sigma_{2x}+\sigma_{1x}\sigma_{2z}$  &   $X_{1}U_{2}$     \\
\hline
 $-\sigma_{1x}-\sigma_{1z}\sigma_{2y}$& $ Y_{1}U_{1}$  &
 $\sigma_{1y}+\sigma_{2z}+\sigma_{1z}\sigma_{2y}-\sigma_{1x}\sigma_{2x}$  & $U_{2}X_{1}Z_{2}$   \\
\hline $-\sigma_{1x}+\sigma_{1z}\sigma_{2z}$&$ Y_{1}U_{1}X_{2}$ &
$\sigma_{1z}-\sigma_{2x}+\sigma_{1x}\sigma_{2z}+\sigma_{1y}\sigma_{2y}$
&$U_{2}Z_{1} Y_{2}$
\end{tabular}
\end{ruledtabular}
\end{table}
\endgroup
The above discussions show that all the single-qubit measurements
can be experimentally implemented by POMs $(|1\rangle\langle
1|)_{l}$ on the given state with or without single-qubit quantum
operations. However, the implementation of multiple-qubit
measurements needs non-local two-qubit operations. The basic
two-qubit operation can be derived from the time-evolution
operator $U_{12}(t)$ of a pair of coupled qubits, labelled by $1$
and $2$, whose Hamiltonian $H_{12}$ can be obtained from
Eq.~(\ref{eq:4}) with $n=2$. Without loss of generality, we
assume: (i)
$\varepsilon_{1\alpha}=\varepsilon_{2\alpha}=\varepsilon_{\alpha}$;
and (ii)
$J_{lm}^{\alpha\beta}=J_{lm}^{\alpha}\delta_{\alpha\beta}$ in the
Hamiltonian~(\ref{eq:4}), because by applying local unitary
operations, e.g.,~\cite{dur}, the Hamiltonian~(\ref{eq:4}) can
always be transformed into a diagonal form---which is actually
used for a number of promising solid-state quantum computing
models~\cite{makhlin,as,loss,ww,kane,dm,zorin,you}. Then in the
basis $\{|0_{1},0_{2}\rangle,
|0_{1},1_{2}\rangle,|1_{1},0_{2}\rangle, |1_{1},1_{2}\rangle\}$,
the time-evolution operator $U_{12}(t)$ is
\begin{equation}\label{eq55}
U_{12}(t)=\exp\left\{-iH_{12}t/\hbar\right\}=
\sum_{g=1}^{4}e^{-iE_{g}t/\hbar}|\psi_{g}\rangle\langle\psi_{g}|
\end{equation}
where $|\psi_{g}\rangle \,\,(g=1,2,3,4)$ are four eigenvectors of
the Hamiltonian $H_{12}$. The corresponding eigenvalues
$E_{1}=-J^x_{12}-J^y_{12}-J^z_{12}$, $E_{2}$, $E_{3}$ and $E_{4}$
are given~\cite{sol} by the solutions of the cubic equation of the
parameter $E$. Here, we only focus on two typical Hamiltonians
which play an important role in the process of two-qubit operation
for the most representative solid state quantum computing models.
One is that all of the one-qubit parameters are switchable, for
example, quantum dots in cavities~\cite{ima}. However, due to
technical constraints and difficulties, it was found~\cite{wu}
that not all the one-qubit parameters are switchable in the
two-qubit operation, for instance, for spin-coupled quantum
dots~\cite{loss}, donor-atom nuclear or electron
spins~\cite{kane}, and  quantum Hall systems~\cite{dm},  two
one-qubit parameters such as $\varepsilon_{x}$ and
$\varepsilon_{y}$ are switchable, but $\varepsilon_{z}$ is fixed.
The basic two-qubit operation with fixed $\varepsilon_{z}$ is
\begin{eqnarray}\label{eq:7}
&&U_{12}(t)=\frac{1}{2}(e^{i\phi}\cos\gamma+e^{-i\phi}\cos\beta )
I+i\frac{(1-a^2)c}{2}e^{-i\phi}\nonumber\\
&&\times
\sin\beta(\sigma_{1z}+\sigma_{2z})+\frac{1}{2}(e^{-i\phi}\cos\beta
-e^{i\phi} \cos\gamma)\,\,\sigma
_{1z}\otimes \sigma _{2z}\nonumber\\
&&-i\frac{1}{2}(e^{i\phi}\sin\gamma +2ac \,e^{-i\phi} \sin\beta)
\,\,\sigma _{1x}\otimes \sigma
_{2x}\nonumber\\
&&-i\frac{1}{2}(e^{i\phi}\sin\gamma -2ac\,
e^{-i\phi}\sin\beta)\,\,\sigma _{1y}\otimes \sigma_{2y}
\end{eqnarray}
where $\gamma=\frac{t}{\hbar }\left(
J_{12}^{x}+J_{12}^{y}\right)$, $\beta=\frac{t}{\hbar
}\sqrt{4\varepsilon^2_{z}+\left( J_{12}^{x}-J_{12}^{y}\right)^2}$,
$\phi=\frac{t}{\hbar }J_{12}^{z}$, $a=2b+\sqrt{4b^2+1}$, with
$b=\varepsilon_{z}/(J^{x}_{12}-J^{y}_{12})$, and $c=1/(1+a^2)$. We
also assume in Eq.~(\ref{eq:7}) that the parameters satisfy
conditions $[2J^{z}_{12}\pm(J^{x}_{12}+J^{y}_{12})]^2\neq
4\varepsilon^{2}_{z}+(J^{x}_{12}-J^{y}_{12})^2$ and
$2J^{z}_{12}\pm\sqrt{4\varepsilon^2_{z}+(J^{x}_{12}-J^{y}_{12})^2}\neq
(J^{x}_{12}+J^{y}_{12})^2$.

{\it Examples of two-qubit measurements.}--- Using
Eq.~(\ref{eq:7}), we can obtain the two-qubit operations by
choosing system-specific parameters. For example, the two-qubit
operation with fixed $\varepsilon_{z}$ for the Heisenberg model,
$XXZ$ model, and the $XY$ model can be obtained from
Eq.~(\ref{eq:7}) by setting parameters
$J^{x}_{mn}=J^{y}_{mn}=J^{z}_{mn}$, $J^{x}_{mn}= J^{y}_{mn}\neq
J^{z}_{mn}$ and $J^{x}_{mn}=J^{y}_{mn}, J^{z}_{mn}=0$,
respectively. If all one-qubit parameters are switchable, then the
two-qubit operation can be obtained from Eq.~(\ref{eq:7}) by only
setting $\varepsilon_{z}=0$. Other effective spin quantum
computing models presented up to now can be reduced by
single-qubit operations to Eq.~(\ref{eq:7}). For instance, i) the
two-qubit operations of the superconducting charge
qubit~\cite{makhlin} can be reduced to Eq.~(\ref{eq:7}) with
$J^{x}_{12}=J^{z}_{12}=0$ by a
conjugation-by-$(\pi/4)(\sigma_{1y}+\sigma_{2y})$~\cite{lidar} on
the Hamiltonian;  ii) the two-qubit operation for the models in
Refs.~\cite{you} and ~\cite{vw} can be reduced to Eq.~(\ref{eq:7})
with $J^{x}_{12}=J^{y}_{12}=0$ and
 model~\cite{makhlin} by the conjugation-by-$(\pi/4) (\sigma_{1y}+\sigma_{2y})$
 and conjugation-by-$(\pi/4) (\sigma_{1x}+\sigma_{2x})$ on the
Hamiltonian of the system. Combining the  basic two-qubit
operations $U_{12}(t)$ with the single-qubit operations, we can
obtain any desired two-qubit operation by choosing the evolution
time $t$ and the system-specific parameters
$\{\varepsilon_{\alpha},J^{\alpha}_{12}:\alpha=x,y,z\}$.

Now, let us consider the $XY$ exchange coupling system with
switchable one-qubit parameters as an example to answer how to
obtain the expectation values of two-qubit measurements. If we
want to obtain, for example, the expectation value $r_{zy}$ of
$\sigma_{1z}\otimes\sigma_{2y}$ in such system, we can first
switch off all the one-qubit parameters $\varepsilon_{l\,\alpha}$,
then let the two-qubit system evolve during the time $\tau=\hbar
\pi/8J^{x}_{12}$ with the evolution operator $U_{12}(\tau)$, then
switch off the exchange coupling and only make the first qubit
have a $\pi/2$ rotation $Y_{1}=\exp\{-i\pi\sigma_{1y}/4\}$ around
the $y$ axis, that is
\begin{equation}
\rho\xrightarrow{U_{12}(\tau)}U_{12}(\tau)\rho\,
U^{\dagger}_{12}(\tau)\xrightarrow{Y_{1}}Y_{1}U_{12}(\tau)\rho
\,U^{ \dagger}_{12}(\tau)Y_{1}^{\dagger}=\widetilde{\rho}.
\end{equation}
Afterwards we make the measurement $(|1\rangle\langle 1|)_{1}$ on
the rotated state $\widetilde{\rho}$ obtaining the probability
$p={\rm Tr}[\widetilde{\rho} \,(|1\rangle\langle 1|)_{1}] =(\sqrt
{2}+r_{x0}+r_{zx})/2\sqrt{2}$,  corresponding to the equivalent
two-qubit measurement $-\sigma_{1x}-\sigma_{1z}\otimes
\sigma_{2x}$. Because $r_{x0}$ has been obtained by the equivalent
single-qubit measurement $\sigma_{1x}$, $r_{zx}$ is completely
determined by the above result. Eight other values of equivalent
two-qubit measurements for this pair can also be obtained by
projecting $(|1\rangle\langle 1|)_{1}$ on the measured state with
the quantum operations summarized in Table~\ref{tab1}. Each
measured value is related to the expectation values of a
single-qubit and a two-qubit measurements. For a two-qubit state
in this system, the above  $9$ two-qubit and $6$ single-qubit
measurements are enough to obtain $15$ unknown parameters
$r_{l_{1}l_{2}} \,(l_{1},l_{2}=0, x,y,z)$ where $l_{1}, l_{2}$ are
not simultaneously taken as zero. The $16$ matrix elements of the
two-qubit state are obtained by combining the $15$ parameters
$r_{l_{1}l_{2}}$ with the normalization condition and finally the
two-qubit state can be completely reconstructed.

The implementation of equivalent two-qubit measurements with a
well-chosen two-qubit operation for a pair of coupled two-qubit
system plays a significant role in the reconstruction of a state.
For the $XY$ and Heisenberg models with switchable one-qubit
parameters, the equivalent measurements
$\sqrt{2}W^{\dagger}\sigma_{1z}W$ and $2W^{\dagger}\sigma_{1z}W$,
to obtain the expectation values of $9$ two-qubit measurements,
are summarized in table~\ref{tab1}, where  the non-local two-qubit
operation operators $U_{1}$ and $U_{2}$ for the $XY$ and
Heisenberg models are chosen by Eq.~(\ref{eq:7}) with the system
specific parameters and the evolution time
$\tau=\hbar\pi/8J^{x}_{12}$ as
\begin{eqnarray}
2\sqrt{2}U_{1}&=&\left(\sqrt{2}+1\right)I
+\left(\sqrt{2}-1\right)\sigma^{1}_{z}\otimes\sigma^{2}_{z}\nonumber\\
&-&i\sigma^{1}_{y}\otimes\sigma^{2}_{y}-i\sigma^{1}_{x}\otimes\sigma^{2}_{x},\\
2\sqrt{2}U_{2}&=&\left(2-i\right)I
-i\sigma^{1}_{z}\otimes\sigma^{2}_{z}-i\sigma^{1}_{y}\otimes\sigma^{2}_{y}\nonumber\\
&-&i\sigma^{1}_{x}\otimes\sigma^{2}_{x}.
\end{eqnarray}
The reconstructions of the qubit states in these models with fixed
$\varepsilon_{z}$ are the same with switchable one-qubit
parameters, if the ratios $\varepsilon_{z}/J^{x}_{12}=4m/(2n-1)$
($m,n=1,2,\cdots$) are appropriately  chosen and the operation
time is $\tau=(n\hbar\pi)/(2\varepsilon_{z})$.

We also find that the tomographic measurement steps for most
systems can be reduced to the same steps needed for the $XY$
model. For example, i) by choosing appropriate values of
$J_{12}^{z}$, $J^{x}_{12}$ ($J_{12}^{z}$, $J^{x}_{12}$, and
$\varepsilon_{z}$) and operation time $\tau$ for the $XXZ$ model
with the switchable one-qubit parameters (fixed $\varepsilon_{z}$)
such that $J^{z}_{12}\tau/\hbar=2n\pi$,
$J^{x}_{12}\tau/\hbar=(2m-1)\pi/8$
 ($J^{z}_{12}\tau/\hbar=2n\pi$, $J^{x}_{12}\tau/\hbar=(2m-1)\pi/8$,
$\varepsilon_{z}\tau/\hbar=l\pi/2$) with $l,m,n=1,2\cdots$, then
we can obtain the same two-qubit operation as for the $XY$ model
and the qubit state can be reconstructed by using the same steps
as the $XY$ model, ii) the qubit state of the superconducting
charge qubit model can also be reconstructed by using the same
steps as the $XY$ model when the parameters  and evolution time
are appropriately chosen~\cite{Liu}, iii) the qubit states in the
systems modelled in Refs~\cite{you} and~\cite{vw} can also be
reconstructed by using the same steps used for the $XY$ model.

It should be emphasized that the different qubit measurements on
the quantum state with fixed quantum operations produce different
results and the quantum operations are not unique to obtain each
expectation value of the measurement. In table~\ref{tab1}, we only
discuss the procedure for the first qubit POM, if we can make all
single-qubit measurements, then the operation steps to obtain some
of the expectation values of the multiple-qubit measurements can
be decreased. For example, if we can experimentally perform the
second qubit projection $(|1\rangle\langle 1|)_{2}$ in the $XY$
model, the expectation value of $\sigma_{1y}\otimes\sigma_{2z}$
can be obtained by using two steps of operations $U_{12}(\tau)$
and $Y_{2}$, that is
$U_{12}^{\dagger}(\tau)Y^{\dagger}_{2}\,\sigma_{2z}
\,Y_{2}U_{12}(\tau)=-(\sigma_{2x}+
\sigma_{1y}\otimes\sigma_{2z})/\sqrt{2}$. But  four steps are
needed for the first qubit measurement.  The price paid is that
noise may be increased because the system is in contact with more
probes.

{\it Multi-qubit measurements.}---These measurements can also be
obtained by designing step-by-step single- and two-qubit
operations. In principle, to obtain the expectation value of an
$l$-qubit $(2<l\leq n)$ measurement, we need at least $l-1$
two-qubit operations for different pairs of the $l$-qubit. For
example, let us obtain the expectation value $r_{zzx}$
corresponding to the three-qubit measurement
$\sigma_{1z}\otimes\sigma_{2z}\otimes\sigma_{3x}$ for the $XY$
interaction system with  switchable one-qubit parameters. We can
replace the first qubit by the third qubit in the two-qubit
operation $U_{12}(\tau)$ and perform an operation $U_{23}(\tau)$
on the second and third qubits, followed by another operation
$U_{12}(\tau)$ on the first and second qubits, followed by a
$\pi/2$ rotation $Y_{1}$ about the $y$ axis for the first qubit,
followed finally by the measurement $(|1\rangle\langle 1|)_{1}$.
That is, an equivalent three-qubit measurement can be obtained as
\begin{eqnarray}
&&U^{\dagger}_{23}(\tau)U^{\dagger}_{12}(\tau)Y_{1}^{\dagger}\sigma_{1z}Y_{1}U_{12}(\tau)U_{23}(\tau)\nonumber\\
&&=-\,\frac{1}{2\sqrt{2}}\sigma_{1x}-\frac{1}{4}\sigma_{1z}\otimes\sigma_{2y}
+\frac{1}{4}\sigma_{1z}\otimes\sigma_{2z}\otimes\sigma_{3x},
\end{eqnarray}
where the assumption of exchange couplings are the same for all
qubit pairs used.  The probability of measuring $(|1\rangle\langle
1|)_{1}$ on the above rotated three-qubit state is
$p^{\prime}=(2\sqrt{2}+2r_{x00}+\sqrt{2}r_{zy0}-\sqrt{2}r_{zzx})/4\sqrt{2}$.
Then $r_{zzx}$ can be determined by the $p$,  $r_{x00}$, and
$r_{zy0}$, the latter two parameters have been obtained by single-
and two-qubit measurements. We can also obtain other probabilities
of the equivalent three-qubit measurements related to the
expectation values of the three-qubit measurements by projecting
$(|1\rangle\langle 1|)_{l}$ on the final operated state. For a
three-qubit state, we can solve the equations for all
probabilities of equivalent one-, two-, and three-qubit
measurements to obtain expectation values of all measurements,
finally all matrix elements of a three-qubit state are obtained by
these expectation values, and the state is reconstructed. Any
$n$-qubit measurement can be obtained in a similar way to the
three-qubit measurement, then the $n$-qubit state can also be
reconstructed.

{\it Discussion.}--- In summary, we have proposed a scheme for
tomographic reconstruction of qubit states for a class of
promising solid-state quantum computing models. We find that
elemental logic gates, such as CNOT gate, control phase gate, etc.
are not necessary in this process. An appropriate non-local
two-qubit operation is enough to realize this purpose. The
generalization of the above discussion to the spin
measurement~\cite{you,zorin} is straightforward because of the
equivalence between $|1\rangle\langle
1|=\frac{1}{2}(\sigma_{0}-\sigma_{z})$ and $\sigma_{x}$
($\sigma_{y}$) by a $\pi/2$ ($-\pi/2$) rotation about the $y$
($x$) axes. Using present technology, our proposal is
experimentally feasible in these solid state qubit systems.
Ideally, the reconstructed qubit state $\rho$ should satisfy the
properties of the normalization, positivity, and Hermiticity.
However, we always deal with a limited amount of experimental
data, which are also affected by noise and imperfect quantum
measurements. To overcome problems due to unavoidable statistical
errors, we can use the maximum-likelihood estimation of the
density matrix~\cite{mj} to obtain a more accurate reconstructed
qubit.

We thank J. Q. You, X. Hu, J. S. Tsai, and  Q. Niu for helpful
discussions. This work was supported in part by the NSA and ARDA
under AFOSR contract No. F49620-02-1-0334, and by the NSF grant
No. EIA-0130383.

\end{document}